\documentclass[sigconf,printcopyright,natbib=false]{acmart}

\usepackage{algorithmic}
\usepackage{graphicx}
\usepackage{textcomp}
\usepackage{xcolor}
\usepackage{makecell}

\AtBeginDocument{%
  \providecommand\BibTeX{{%
    \normalfont B\kern-0.5em{\scshape i\kern-0.25em b}\kern-0.8em\TeX}}}

\copyrightyear{2022}
\acmYear{2022}
\setcopyright{acmlicensed}\acmConference[ASE '22]{37th IEEE/ACM International Conference on Automated Software Engineering}{October 10--14, 2022}{Rochester, MI, USA}
\acmBooktitle{37th IEEE/ACM International Conference on Automated Software Engineering (ASE '22), October 10--14, 2022, Rochester, MI, USA}
\acmPrice{15.00}
\acmDOI{10.1145/3551349.3559573}
\acmISBN{978-1-4503-9475-8/22/10}

\begin{document}

\title{Call Graph Evolution Analytics over a Version Series of an Evolving Software System}

\author{Animesh Chaturvedi}

\orcid{0002-9058-9052}
\affiliation{%
  \institution{Indian Institute of Information Technology (IIIT), Dharwad}
  \streetaddress{P.O. Box 1212}
  \city{Dharwad}
  \state{Karnataka}
  \country{India}}
\email{animesh.chaturvedi88@gmail.com}

\begin{abstract}
  Software evolution analytics can be supported by generating and comparing call graph evolution information over versions of a software system. Call Graph evolution analytics can assist a software engineer when maintaining or evolving a software system. This paper proposes \textit{Call Graph Evolution Analytics} to extract information from a set of \textit{Evolving Call Graphs} ECG = $\{CG_1, CG_2,... CG_N\}$ representing a Version Series VS = $\{V_1, V_2, ... V_N\}$ of an evolving software system. This is done using \textit{Call Graph Evolution Rules} (CGERs) and \textit{Call Graph Evolution Subgraphs} (CGESs). Similar to association rule mining, the CGERs are used to capture co-occurrences of dependencies in the system. Like subgraph patterns in a call graph, the CGESs are used to capture evolution of dependency patterns in evolving call graphs. Call graph analytics on the evolution in these patterns can identify potentially affected dependencies (or procedure calls) that need attention. The experiments are done on the evolving call graphs of 10 large evolving systems to support dependency evolution management. This is demonstrated with detailed results for evolving call graphs of Maven-Core's version series.
\end{abstract}



\maketitle

\section{Introduction}
Four phases in a software project includes: requirements gathering, development, testing, and maintenance/evolution \cite{Rajlich2014}. In the maintenance phase, a software repository holds plenty of information about the dependency evolution in call graphs. Software repositories are of various kinds; this study focuses on dependency repositories (e.g. Maven and GitHub), which contain jars, APIs, and libraries. Source code or executables are retrieved from a repository, and used in a software project. Growth of software - in the numbers of repositories and in the size of repositories - makes mining of a dependency repository a challenge. Existing techniques for mining software repositories \cite{Hassan2008} and search based software engineering \cite{Harman2007} commonly analyse a single software version. 

A call graph \cite{Ryder1979}\cite{Grove2001} represents procedure calls (or dependency) relationships in the form of a directed graph, where nodes represent procedures (method or function) and directed edges represent dependencies from caller to callee procedures. Earlier call graph analysis techniques have proven to be helpful in representing software component interactions \cite{Hashim2012}, program comprehension of large software systems \cite{Shaw2002}, impact analysis based on mutation testing \cite{Musco2017}, auto-completion of the procedure calls \cite{Garbervetsky2017}, and anomaly detection for monitoring system calls \cite{Gao2004}. Graph or network theory includes e.g., temporal networks \cite{Holme2005}, evolving networks \cite{Aggarwal2014}, dynamic networks \cite{Aronson1989}, and multilayer networks \cite{Domenico2015}. The evolution of procedure calls can be exploited using techniques such as change mining \cite{Mirko2008} and evolution mining \cite{Mansoureh2011}\cite{Bin2007}.

This paper proposes two forms of Call Graph Evolution mining techniques to uncover interesting and useful information: Call Graph Evolution Rules (CGERs) and Call Graph Evolution Subgraphs (CGESs). For software evolution analytics, apply graph evolution mining on multiple call graphs representing multiple versions of a software repository. Use call graph evolution that provides information about the software evolution. Call graph evolution information enables tools to support and manage the evolution of dependencies. Section II presents an abstract of the two proposed techniques. Section III demonstrates results (Stable CGERs and CGESs), and also discusses a \textbf{Research Question} “How does the results correspond to the Lehman's law of software evolution?”. 

\section{Call Graph Evolution Analytics}
In an evolving software, each version (say $V_i$) of the software has its call graph, which can be built by capturing the procedure call relationship between the dependencies of version $V_i$. This pre-processing involves various steps (e.g., parsing) that depend upon the evolving software. The number of procedures is constant for a version, however different versions have different numbers of procedures over time. The series of software versions can be represented as a series of call graphs, which is referred to as Evolving Call Graphs. Here, ``Evolving'' denotes call graphs belonging to multiple versions of an evolving software. A series of call graphs are Evolving Call Graphs ECG = $\{CG_1, CG_2,... CG_N\}$ that represents a Version Series VS = $\{V_1, V_2, ... V_N\}$ of an evolving software.

On a fine-grained level, the rule mining retrieves co-occurrences of dependencies in a call graph of a version, and then counts stability of retrieved rules over the version series. On a coarse-grained level, the subgraph mining extracts frequently occuring structural patterns of dependencies in a call graph of a version, and then aggregates frequencies of retrieved patterns over the version series.

\subsection{Stable Call Graph Evolution Rules (CGERs)}
\begin{figure}[htbp]
\includegraphics[width=1.0 \linewidth]{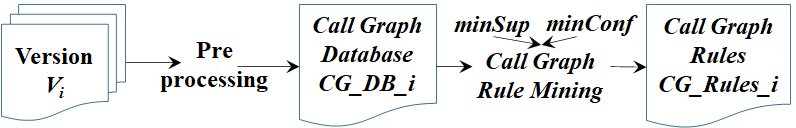}
\caption{Call Graph Rule mining on version ($V_i$).}
\label{Fig1}
\end{figure}

Initially, perform Call Graph Rule (CGR) mining for each version of an evolving software in two steps. First, given a version of the system, the procedures of the system are assumed to be divided up into modules, which is referred to as the “procedures-module membership information”. Transform this information into a set of “call pairs”, which form the core or the Call Graph Database CG\_Db. Then, pre-process a version $V_i$ to create a call graph database (CG\_Db\_i) as shown in Fig. \ref{Fig1}. Second step uses this database with two rule mining thresholds, minimum support (minSup) and minimum confidence (minConf), to generate a collection of Call Graph Rules (CG\_Rules\_i) as shown in Fig. \ref{Fig1}. The Call Graph Rule X $\rightarrow$ Y can be interpreted as “if the procedure(s) of set X appear in a call pair, then the procedure(s) of set Y are likely to appear in the module with support and confidence above minSup and minConf, respectively”. In other words, the presence of the caller procedure(s) in X implies the presence of the callee procedure(s) in Y with sufficiently high frequency to surpass the two thresholds minSup and minConf. This formally defines the Call Graph Rule with its support and confidence, and when a rule is interesting in a version $V_i$.

\begin{figure}
\centering
\includegraphics[width=0.5 \linewidth]{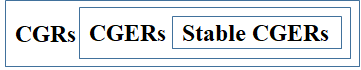}
\caption{Venn diagram of CGRs, CGERs, and Stable CGERs.}
\label{Fig2}
\end{figure}

\begin{figure}[htbp]
\includegraphics[width=1.0 \linewidth]{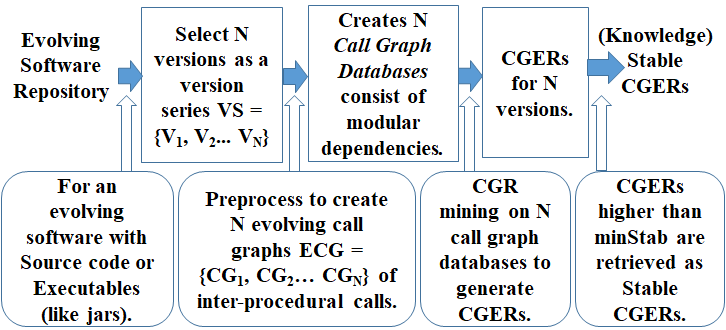}
\caption{Overview of the Stable CGER mining steps.}
\label{Fig3}
\end{figure}

Next, defining the Call Graph Evolution Rules (CGERs) mining over a version series. Out of the retrieved CGRs, select a collection of distinct CGRs as CGERs. Out of the CGERs, retrieve Stable CGERs. Then, there will be a subset relationship between CGRs, CGERs, and SCGERs as shown in Fig. \ref{Fig2}. The CGR is a rule derived from a call graph of a software version. The CGER is a `distinct' CGR from a collection of CGRs in multiple versions. Count of a reoccurring CGR in multiple versions depicts its stability over a version series. The CGERs accompany the count (as stability). Each CGER has its `stability', which is the count of versions in which CGR appears interesting (i.e., its support and confidence is above minSup and minConf). Earlier, we defined minimum Stability (minStab) \cite{Chaturvedi2019minStab}\cite{Chaturvedi2022SysNetAnalytics}. A CGER is defined as a Stable CGER in VS, if (a) the support and confidence are greater than minSup and minConf in a version $V_i$, and (b) the `stability' ($\geq$) minStab (i.e. stability of CGERs is greater than the minStab). This means a CGER is stable in VS if it exceeds the three user-specified thresholds: minSup, minConf, and minStab. These CGERs can also be referred to as call graph prediction rules (temporal rules, or episode rules). This takes the general form “if a certain dependency(ies) occurs, then another dependency(ies) is(are) likely to occur in the module”. The Stable CGERs Mining steps are shown in Fig. \ref{Fig3}.

\subsection{Call Graph Evolution Subgraphs (CGES)}
\begin{figure}
\includegraphics[width=1.0 \linewidth]{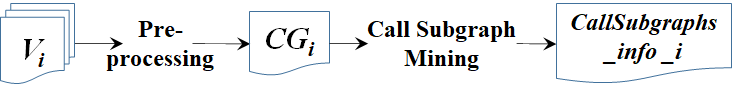}
\caption{Call Subgraph Mining for Version $V_i$.}
\label{Fig4}
\end{figure}

The CGES mining retrieves the evolution of subgraphs in call graphs. Software version $V_i$ is first pre-processed to build its call graph, which is used as the input for subgraph mining to retrieve subgraph patterns with their frequencies as shown in Fig. \ref{Fig4}. Apply subgraph mining to each of the call graphs in a set of N call graphs. This iteratively retrieves the graphlets information for all versions of a software to extract the Call Graph Evolution Subgraphs and their frequencies. These subgraphs are of two types \textit{Call Graph Evolution Graphlets} (CGEGs) and \textit{Call Graph Evolution Motifs} (CGEMs).

\begin{figure}[htbp]
\includegraphics[width=1.0 \linewidth]{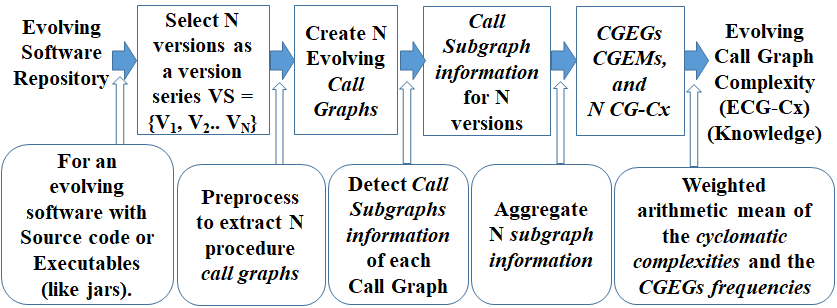}
\caption{Steps of the CGES mining.}
\label{Fig5}
\end{figure}

Further, the Call Graph Evolution Graphlet information (subgraphs and their frequencies) are used to calculate a Call Graph Complexity (CG-Cx) of a single version and the overall Evolving Call Graph Complexity (ECG-Cx) for a version series. The Fig. \ref{Fig5} shows the steps to detect CGEGs, CGEMs, CG-Cx, and ECG-Cx by following the guidelines of knowledge discovery and data-mining (KDD). Our approach discovers hidden dependency evolution patterns, which is captured in CGEGs and CGEMs. The frequencies of graphlets are meaningful quantities that are used to calculate complexity for a call graph.

\section{Evolving Call Graphs Analytics}

This section presents study on 10 large evolving software given in the Table \ref{tab1}. The table provides the version series used to perform experiments, number of procedures in those versions, and average number of neighbours of each procedure. Firstly, for each one of them, make a series of evolving call graphs. Secondly, apply the call graph evolution analytics based on the two techniques: Stable CGERs mining and CGESs mining. These two techniques retrieved CGRs, CGERs, Stable CGERs, CGEGs, and CGEMs. The experimentation results of the two techniques are further used to study the persistence and complexity of dependencies in evolving call graphs. Out of ten evolving software systems, detailed demonstration of results for one evolving software i.e., Maven-Core is presented.

\begin{table}[htbp]
\caption{Information about Evolving Call Graphs}
\begin{center}
\begin{tabular}{|c|c|c|c|}
\hline
\textbf{Evolving}&\multicolumn{3}{|c|}{\textbf{Evolving Call Graphs}} \\
\cline{2-4} 
\textbf{software} & Version series & \makecell{\# proce- \\ dures} & \makecell{Average \# \\ neighbours} \\
\hline
\makecell{Commons \\ Codec} & \makecell{\{1.1, 1.2, 1.3, 1.4, 1.5, \\ 1.6, 1.7, 1.8, 1.9, 1.10\}}
& 162 & 1.995 \\
\hline
Guava & \makecell{\{12, 13, 14, 15, \\ 16, 17, 18, 19\}} & 1281 & 2.471 \\
\hline
\makecell{Hadoop \\ HDFS} & \makecell{\{2.2.0, 2.3.0, 2.4.0, 2.4.1, \\ 2.5.0, 2.5.1, 2.5.2, 2.6.0, \\ 2.6.1, 2.6.2, 2.6.3, 2.6.4, \\ 2.7.0, 2.7.1, 2.7.2\}}
& 3129 & 2.166 \\
\hline
\makecell{HTTP \\ Client} & \makecell{\{4.3.1, 4.3.2, 4.3.3, 4.3.4, \\ 4.3.5, 4.3.6, 4.3.0, 4.4.1, \\ 4.4.0, 4.5.1, 4.5.2, 4.5.0\}} & 276 & 2.020 \\
\hline
\makecell{JMeter \\ Core} & \makecell{\{1.8.1, 1.9.1, 2.0, 2.1, 2.2, \\ 2.3, 2.4, 2.5, 2.6, 2.7, 2.8, \\ 2.9, 2.10, 2.11, 2.12, 2.13 \}} & 806 & 2.632 \\
\hline
\makecell{Joda \\ Time} & \makecell{\{2.0, 2.1, 2.2, 2.3, 2.4, 2.5, \\ 2.6, 2.7, 2.8.0, 2.8.1, 2.8.2, \\ 2.9.0, 2.9.1, 2.9.2, 2.9.3\}} & 418 & 2.886 \\
\hline
JUnit & \makecell{\{4.1, 4.6, 4.7, 4.8, \\ 4.9, 4.11, 4.12 \}} & 171 & 1.628 \\
\hline
Log4J & \makecell{\{1.1.3, 1.2.4, 1.2.5, 1.2.6, \\ 1.2.7, 1.2.8, 1.2.9, 1.2.11, \\ 1.2.12, 1.2.13, 1.2.14, \\ 1.2.15, 1.2.16, 1.2.17\}} & 307 & 2.320 \\
\hline
\makecell{Maven \\ Core} & \makecell{\{3.1.0, 3.1.1, 3.2.1, \\ 3.2.2, 3.2.3, 3.2.5, \\ 3.3.1, 3.3.3, 3.3.9\}} & 594 & 2.385 \\
\hline
\makecell{Storm \\ Core} & \makecell{\{0.9.1, 0.9.2, 0.9.3, \\ 0.9.4, 0.9.5, 0.9.6, \\ 0.10.0, 0.10.1\}} & 373 & 1.686 \\
\hline
\multicolumn{4}{l}{${\#}$ Number of}
\end{tabular}
\label{tab1}
\end{center}
\end{table}

\begin{figure}
\includegraphics[width=1.0 \linewidth]{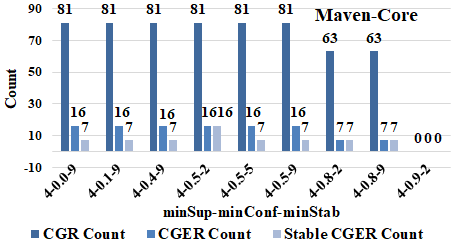}
\caption{Compare counts of CGRs, CGERs, \& Stable CGERs.}
\label{Fig6}
\end{figure}

\begin{figure*}
\includegraphics[width=1.0 \linewidth]{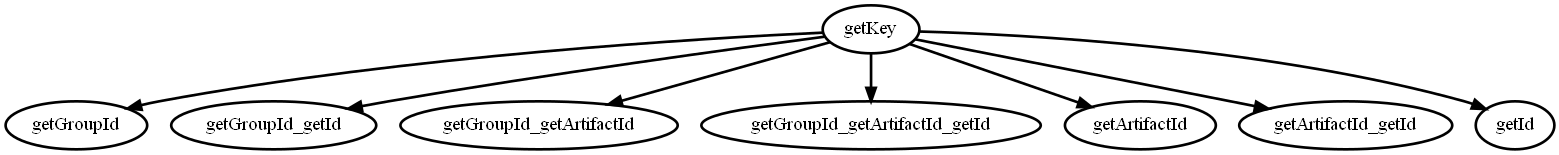}\\
\includegraphics[width=0.8 \linewidth]{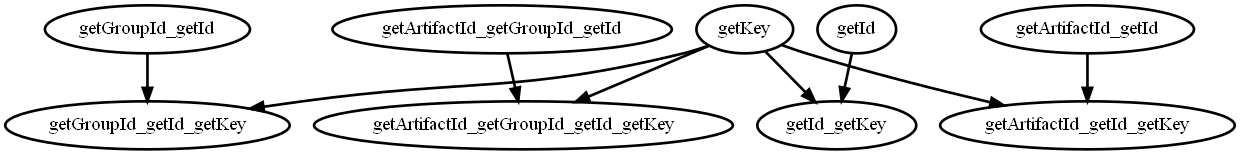}
\vspace{-4pt}
\caption{Transitivity and Lattice graphs respectively for given Stable CGERs in Table 2 for the Maven-Core (evolving software).}
\vspace{-7pt}
\label{Fig7}
\end{figure*}

- The Fig. \ref{Fig6} shows the comparison between the number of CGRs, CGERs, and Stable CGERs for Maven-Core. The figure shows nine experiments, in each experiment we identified that interesting CGERs and Stable CGRs are fewer than CGRs. The 16 Stable CGERs retrieved for the Maven-Core are given in the first column of Table \ref{tab2}. Some of the transitivities and lattices formed based on these Stable CGERs are shown in the Fig. \ref{Fig7}, which shows relationships between the various procedures in the Maven-Core.

\begin{figure}
\includegraphics[width=1.0 \linewidth]{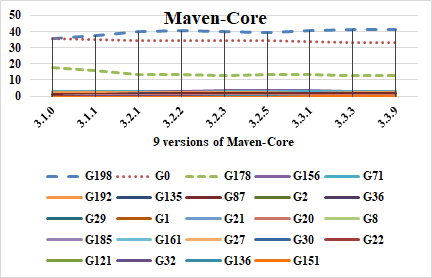}
\vspace{-15pt}
\caption{Frequencies of various CGEGs of evolving call graphs representing the version series of Maven-core.}
\label{Fig8}
\vspace{-8pt}
\end{figure}

\begin{figure}
\includegraphics[width=1.0 \linewidth]{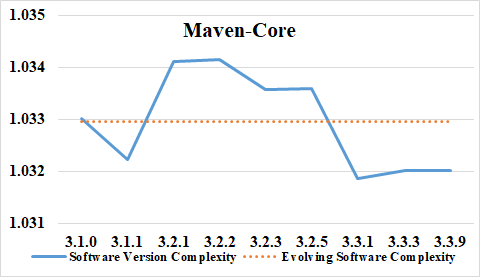}
\vspace{-15pt}
\caption{The varying complexities for each version and a constant aggregated complexity over multiple versions.}
\label{Fig9}
\vspace{-8pt}
\end{figure}

- Various CGEGs are retrieved from the evolving call graphs of the Maven-core as shown in the Fig. \ref{Fig8}, which shows the CGEGs with their frequencies over a version series. These frequencies of CGEGs along with their cyclomatic complexities are used to retrieve varying complexity of each call graph (shown with a varying time-series in Fig. \ref{Fig9}) and an aggregated complexity of all evolving call graphs (shown with a constant time-series in Fig. \ref{Fig9}). Out of many CGEGs, few statistically significant CGEMs patterns with their frequencies above a certain threshold are shown in the second column of Table \ref{tab2}.

\begin{table}[htbp]
\caption{Experimental results on Maven-Core.}
\begin{center}
\begin{tabular}{|c|c|}
\hline
\textbf{Stable CGERs} & \textbf{CGEMs} \\
\hline

\makecell{\{getGroupId $\rightarrow$ getArtifactId\} \\ \{getGroupId $\rightarrow$ getArtifactId, getId\} \\ \{getArtifactId $\rightarrow$ getGroupId\} \\ \{getArtifactId $\rightarrow$ getGroupId, getId\} \\ \{getGroupId $\rightarrow$ getId\} \\ \{getId $\rightarrow$ getGroupId\} \\ \{getId $\rightarrow$ getGroupId, getArtifactId\} \\ \{getKey $\rightarrow$ getGroupId\} \\ \{getKey $\rightarrow$ getGroupId, getId\} \\ \{getKey $\rightarrow$ getGroupId, getArtifactId\} \\ \{getKey $\rightarrow$ getGroupId, getArtifactId, getId\} \\ \{getArtifactId $\rightarrow$ getId\} \\ \{getId $\rightarrow$ getArtifactId\} \\ \{getKey $\rightarrow$ getArtifactId\} \\ \{getKey $\rightarrow$ getArtifactId, getId\} \\ \{getKey $\rightarrow$ getId\}} & \makecell{
\includegraphics[width=0.15 \linewidth]{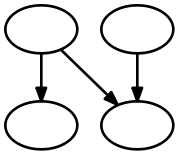}
\\
($M_{198}$, 39.55)
\\
\includegraphics[width=0.2 \linewidth]{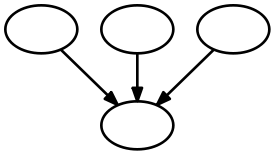}
\\
($M_0$, 34.25)
\\
\includegraphics[width=0.2 \linewidth]{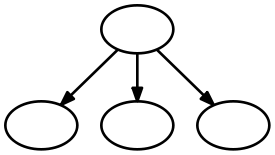}
\\
($M_{178}$, 13.85)
} \\

\hline
\end{tabular}
\label{tab2}
\end{center}
\end{table}

The two kinds of experiments to retrieve Stable CGERs and CGESs made us infer that changes in an evolving software lead to several versions, whose dependency patterns also evolve in the versioning process to represent a state of the evolution information. Then, aggregate the state information over a version series to achieve evolution information about a whole centralized repository. This evolution information is helpful to manage and track evolution happening in a version series of an evolving software.

\textbf{Answer to Research Question:} For the call graph evolution, we found three inferences with Lehman's law of software evolution (in 1980) \cite{Lehman1980}. We found two similar inferences. An evolving software undergoes continuous upgrading to make better versions (in Table I). The stability of an evolving software remains almost constant with time (in Fig. \ref{Fig8}). We found one dissimilar inference. The complexity of evolving software keeps on changing with time, but not necessarily increasing (in Fig. \ref{Fig9}). This is because, now there are better software repository management techniques and systems (e.g. GitHub, Maven) are easily accessible as compared to the era when Lehman's law was introduced. 

Like Vasa and Schneider \cite{Vasa2003} found patterns having low cyclomatic complexity occur with high percentage in a version. We also found recurring subgraph patterns with low complexity occur in high percentage over a version series. These subgraphs are the CGEMs \{$M_{198}$, $M_0$, $M_{178}$\} with cyclomatic complexity = 1 in Table \ref{tab2}, which has high frequencies of CGEGs \{$G_{198}$, $G_0$, $G_{178}$\} in Fig. 8.

\vspace{-1pt}
\section{Related works and discussions}
\vspace{-1pt}
On one hand, 1970-2010 was the era of control flow graph analysis \cite{Allen1970} and dependency analysis \cite{Sharma2009}, which are exploited to do the software analysis \cite{Jackson2000}, program slicing \cite{Weiser1984}\cite{Binkley1996}, and source code analysis \cite{Binkley2007}. On the other hand, in 1993 association rule mining proposed by Agrawal et al. \cite{Agrawal1993}, which was used in the context of software engineering. Ying Annie et. al. \cite{Ying2004} (in 2004) predicted a set of possible future dependencies across files using support metrics to mine history of software changes. Zimmerman et al. \cite{Zimmermann2005} (in 2005) applied association rule mining on software version histories to predict and suggest changes, co-changes, and prevent errors. Works like \cite{Ying2004}\cite{Zimmermann2005} made the field of Mining Software Repositories.

Vasa et al. \cite{Vasa2007} presented multiple releases of object-oriented classes and interfaces for stability and complexity. Pang et al. \cite{Pang2015} proposed N-gram analysis and prediction of vulnerable components for features like sequences in source code files, and Java class files. Kikas et al. \cite{Kikas2017} discussed the structure and evolution of dependency graphs collected from the package. Honsel \cite{Honsel2015} retrieved  interesting evolution patterns in the dependency graphs. Decan et al. \cite{Decan2019} presented empirical analysis of Dependency Network Evolution on seven packages. In comparison to these works, this paper presented a novel way to use call graphs evolution mining for studying structural patterns, stability, and complexity.

\vspace{-1pt}
\section{Conclusions}
\vspace{-1pt}
This paper introduced two proposed techniques, which are used to analyze 10 evolving software systems (including Maven-Core described in detail). Looking forward to publishing a complete study of Stable CGERs mining and CGESs mining, which will provide intrinsic details of the approaches and demonstrate detailed study of 10 evolving software systems. Other related works by the author on System Evolution Analytics for Hadoop-HDFS repository \cite{Chaturvedi2019minStab}-\cite{Chaturvedi2022SysNN} and Cloud Service Evolution Analytics \cite{Chaturvedi2014Subset}-\cite{Chaturvedi2022Service}. 

\vspace{-1pt}
\begin{acks}
\vspace{-1pt}
Thanks to Prof. Aruna Tiwari, Prof. Dave Binkley, and Prof. Nicolas Spyratos for advice and feedback to make the author a PhD graduate from Indian Institute of Technology Indore (IIT Indore).
\end{acks}

\end{document}